\documentstyle[natbib2, epsfig,natbibmnfix,epstopdf,amsmath]{mn2e}

\def\msun{{\,{\rm M}_\odot}}

\def\simlt{\lower.5ex\hbox{$\; \buildrel < \over \sim \;$}}
\def\simgt{\lower.5ex\hbox{$\; \buildrel > \over \sim \;$}}

\title{Massive black hole binary mergers within sub-pc scale gas discs} 

\author[J.\ Cuadra et al]{J.\ Cuadra$^{1}$\thanks{
Present address: Shanghai Astronomical Observatory,
Shanghai 200030, China; 
jcuadra@shao.ac.cn}, 
P.\ J.\ Armitage$^{1,2}$,
 R.\ D.\ Alexander$^{3,1}$, M.\ C.\ Begelman$^{1,2}$
\\
$^{1}$JILA, University of Colorado and National Institute of Standards and Technology, Boulder, CO 80309-0440, USA
\\
$^{2}$Department of Astrophysical and Planetary Sciences, University of Colorado, Boulder CO 80309, USA \\
$^{3}$Leiden Observatory, Universiteit Leiden, Niels Bohrweg 2, 2300 RA, Leiden, the Netherlands
}

\begin{document}

\date{Accepted XXX. Received XXX; in original form XXX}

\pagerange{\pageref{firstpage}--\pageref{lastpage}} \pubyear{2008}

\maketitle

\label{firstpage}

\begin{abstract}
We study the efficiency and dynamics of supermassive black hole binary mergers  
driven by angular momentum loss to small-scale gas discs. Such binaries form 
after major galaxy mergers, but their fate is unclear since hardening through stellar scattering
becomes very inefficient at sub-parsec distances. Gas discs may dominate 
binary dynamics on these scales, and promote mergers. Using numerical simulations, 
we investigate the evolution of the semi-major axis and eccentricity of binaries 
embedded within geometrically thin gas discs. Our simulations directly resolve 
angular momentum transport within the disc, which at the radii of interest is 
likely dominated by disc self-gravity. We show that the binary decays
at a rate which is in good agreement with analytical estimates, while the 
eccentricity grows. Saturation of eccentricity growth is not observed up 
to values $e \simgt 0.35$. Accretion onto the black holes
is variable, and is roughly modulated by the binary orbital frequency. Scaling  
our results, we analytically estimate the maximum rate of binary decay that 
is possible without fragmentation occuring within the surrounding gas disc, 
and compare that rate to an estimate of the stellar dynamical hardening rate. 
For binary masses in the range $10^5 \msun \simlt M \simlt 10^8 \msun$ we 
find that decay due to gas discs may dominate for separations below  
$a \sim 10^{-2} \ {\rm pc} - 0.1 \ {\rm pc}$, in the regime where the 
disc is optically thick. The minimum merger time scale is shorter than the 
Hubble time for $M \simlt 10^7 \msun$. This implies that gas discs could 
commonly attend relatively low mass black hole mergers, and that a significant 
population of binaries might exist at separations of a few hundredths of a 
pc, where the combined decay rate is slowest. For more massive binaries, where this mechanism fails to act quickly 
enough, we suggest that scattering of stars formed within a fragmenting 
gas disc could act as a significant additional sink of binary angular momentum.
\end{abstract}


\section{Introduction}
The assembly of present-day galaxies occurs via the hierarchical merger of lower mass 
progenitors, many of which seem likely to have supermassive black holes (SMBHs) in 
their nuclei. Following a merger of two galaxies, each harbouring a black hole, we 
expect the SMBHs to sink toward the centre of the merger product and eventually 
form a black hole binary. What happens subsequently is less clear. There is no doubt that 
at sufficiently small separations ($a \simlt 10^{-3} \ {\rm pc}$) loss of angular momentum 
to gravitational radiation occurs rapidly enough to effect coalescence. However, at larger separations 
an as-yet undetermined mix of angular momentum loss to stars and to gas is required 
to bring the black holes into the gravitational radiation inspiral regime. Observations 
provide scant information. Detecting black hole binaries on pc or sub-pc scales (where the black holes' 
gravity dominates the galactic potential) is extremely difficult, with the closest confirmed 
binary in the radio galaxy 0402+379 \citep{Rodriguez06} having a projected separation 
of 7.3~pc. A variety of circumstantial evidence -- of which the most 
compelling is probably the small observed scatter in the relationships between 
black hole mass and galaxy properties \citep{Ferrarese06,Gebhardt00} -- suggests 
that binaries do merge, though these arguments do not constrain the time scale 
or the mechanism that facilitates mergers.

One longstanding motivation for considering the role of gas in
hastening binary mergers has come from the fact that stellar dynamical
processes may simply fail. Although dynamical friction against the
stellar background is rapid for large separations
\citep{Milosavljevic03}, for pc-scale binaries the rate slows
dramatically as the black holes eject the finite number of stars that
have orbits that come close enough to the centre to interact with the
binary. Replenishment of stars on these loss cone orbits occurs on the
relaxation time, which typically exceeds the Hubble time
\citep{Begelman80,Yu02b,Merritt06}. This `last parsec' problem occurs
for spherical or axisymmetric nuclei, and is particularly severe for
massive galaxies which are observed to possess rather low density
stellar cores \citep{Faber97}.  More recently, interest in this long
standing problem has been further piqued by the realization that {\em
  if} gas is responsible for driving mergers, then it is likely that
some of that material will survive in the immediate vicinity of the
holes at the moment of coalescence. Plausibly, some small fraction of
the energy released during the merger may go into heating the gas,
producing an electromagnetic precursor \citep{Armitage02} or afterglow
\citep{Milosavljevic05,Lippai08,Schmittman08,Shields08} to the
gravitational wave event. Detection of an electromagnetic counterpart
would greatly improve the accuracy with which space-based
gravitational wave observatories such as {\em LISA} can localize
mergers, and allow much more information to be gathered about the
properties of the host galaxy \citep{Kocsis07}. It may also be possible to
determine whether gas discs typically attend mergers by
searching for periodic electromagnetic signals
produced by binaries at larger separations,
independent of any gravitational wave information
\cite{Haiman08}.

Theoretically, it is well established that the dissipative nature of
gas allows inflow down to pc scales in galactic nuclei following
galactic mergers
\citep{Mihos96,DiMatteo05b,Mayer07b,Levine08}. Perhaps more
surprisingly, smaller amounts of gas appear to be able to penetrate
deep within the black hole's sphere of influence even in galaxies such
as the Milky Way that are far removed from any significant merger
activity. Evidence for such inflows comes from observations of young
stars with disc-like kinematics close to Sgr~A*
\citep{Paumard06,Lu08}, which may well have formed in situ from gas at
sub-pc scales \citep{Levin03,NC05}. Taken together, these lines of
argument suggest that it is plausible to invoke the existence of gas
discs at small radii as a common feature of galactic nuclei, though
the typical masses and structure of such discs are subject to
considerable uncertainty. Prior theoretical work has established that
gas discs are likely to be important for SMBH binary mergers, provided
that the mass of gas in the disc is roughly of the same order of
magnitude as the mass in black holes
\citep{Gould00,Armitage02,Escala05,Dotti07,MacFadyen08,Hayasaki08}.

In this paper we revisit the interaction between a SMBH binary and a relatively 
low mass circumbinary gas disc. Our main innovation is to study this interaction 
using numerical simulations that directly resolve the physical mechanism of 
angular momentum transport within the gas. At the radii of greatest interest for the SMBH binary 
merger problem (typically tenths of a pc, where the stellar dynamical time scales 
are longest for many galaxies) it is expected that self-gravity will dominate the 
dynamics of surrounding gas discs \citep{Shlosman90,Goodman03}. The onset 
of self-gravity in an accretion disc can result either in a quasi-stable disc with 
outward angular momentum transport, or in fragmentation of the disc into stars. 
Either possibility is of great interest for its effect of SMBH binary mergers. Here, 
we study directly the first regime in which the disc remains stable and angular 
momentum is transported by the action of self-gravitating spiral arms. Physically 
this occurs if the cooling time in the disc exceeds the local dynamical time 
\citep{Gammie01}. We use the simulations to study the rate at which the gas 
drives the binary toward merger, and the effect that the gas disc has on the 
eccentricity of the binary. Previous simulations that have used either artificial 
viscosity or a Navier-Stokes formulation\footnote{Although the use of a 
Navier-Stokes viscosity is an improvement over a reliance on purely 
numerical effects to transport angular momentum, it too may lead to 
unphysical behaviour. In particular, it is known that a disc subject to only 
a Navier-Stokes shear viscosity is unstable to the growth of eccentricity 
even in the absence of perturbations \citep{Ogilvie01}. In general, there 
is no reason to assume that a `turbulent viscosity' resulting from a 
physical process such as self-gravity or the magnetorotational instability 
\citep{Balbus98,Balbus99} will behave in the same way as a microscopic fluid 
viscosity.} to model angular momentum transport 
have found that eccentricity growth (both of the binary, and within the gas disc) 
attends the decay of the binary semi-major axis \citep{Artymowicz91,Armitage05,MacFadyen08}.
The time dependent accretion signatures that result may permit the identification 
of black hole binaries within gas rich systems prior to the final coalescence phase.

\begin{figure*}
   \centerline{\epsfig{file=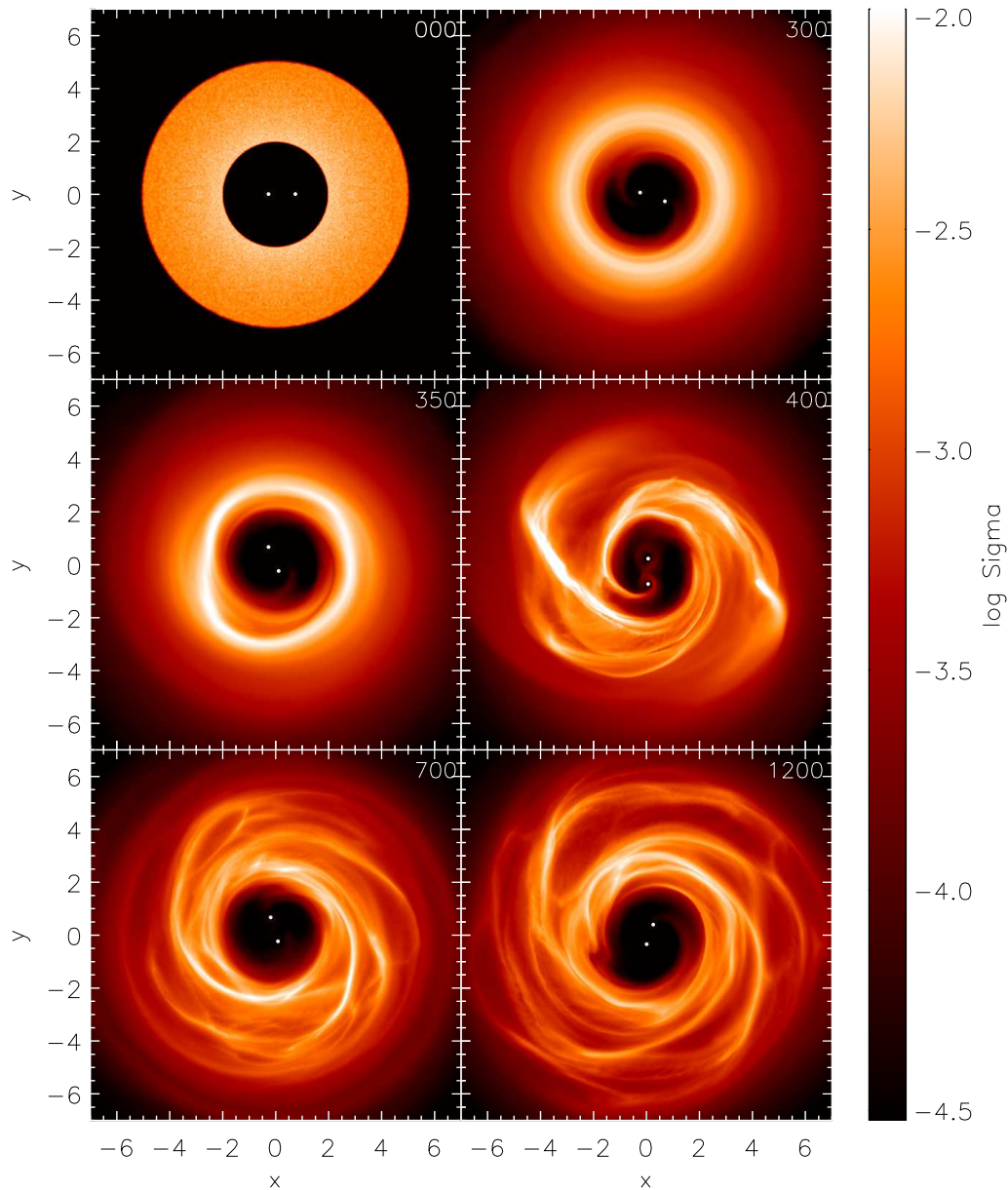,width=.95\textwidth}} 
   \caption{Logarithmic maps of the disc column density (in units of $M a_0^{-2}$) at different times during the simulation.  The panel at $t=0$ shows the smooth initial conditions.  Until $t \approx 350\Omega_0^{-1}$, material piles up at $R \approx 3 a_0$ as a result of the torques shown in Fig.~\ref{fig:torques}, forming a dense ring.  The ring breaks due to its self-gravity (see Fig.~\ref{fig:toomre}), spreading the gas approximately over the original radial range.  A spiral pattern develops, and the disc stays in that state until at least $t \approx 1200\Omega_0^{-1}$, when the simulation ends.}
   \label{fig:disc6}
\end{figure*}

\section{Initial conditions and numerical method}
\label{sec:ics}

In this paper we present simulations that follow the evolution of a binary that initially has a total mass $M$, separation $a_0$, and orbital
frequency $\Omega_0 = \sqrt{GM/a_0^3}$.  The mass ratio between the two components is set to $q = M_1/M_2 = 3$, as typically expected for a
major galaxy merger \citep{Volonteri03}.  We try two different initial orbital configurations for the binary: a circular orbit of radius $a_0$
and an eccentric orbit with semi-major axis size $a_0$ and eccentricity $e=0.1$. 

Around the binary we set a co-rotating circular disc of gas. The disc is aligned with the orbital angular momentum of the  binary, has a total
mass of $M_{\rm d} = 0.2M$ and extends initially from $2a_0$ to $5a_0$.  The surface density of the disc decreases with radius $R$ (measured
from the binary centre of mass) as $\Sigma(R) \propto R^{-1}$.  The disc dimensionless thickness is initially constant $H/R = 2 M_{\rm d} / M$. 
We set the initial internal energy of the gas accordingly as $u = (H/R)^2 v_{\rm K}^2 / (\gamma(\gamma-1))$, where $v_{\rm K} = \sqrt{GM/R}$ is
the Keplerian velocity around the binary and $\gamma=5/3$ is the adiabatic index of the gas.  The disc stability to self-gravity is measured by
the Toomre parameter $Q = \Omega c_{\rm s} / (\pi G \Sigma)$, which translates to  $Q \approx (H/R) (M/M_{\rm d})$ for a disc in hydrostatic
equilibrium.  In our case, the thickness of the disc ensures that the disc is not {\it initially} unstable to self-gravity,
as the Toomre parameter is $\approx 2$.  Only after cooling has affected the gas, will the disc get thinner and become unstable.

To calculate the evolution of the system we use the smoothed particle hydrodynamics \citep[SPH; e.g.,][]{Monaghan92} code {\sc Gadget-2}
\citep{Springel05b}.  The code solves for the adiabatic hydrodynamics and gravitational forces of the system, and includes a term for artificial
viscosity, necessary to treat shocks.  

On top of the basic hydrodynamics set-up we add cooling.  This is implemented by defining a cooling time and setting the
radiative cooling term for each gas particle to be $(du/dt) _{\rm cooling} = -u/t_{\rm cool}$.  The cooling time is set to be proportional to the
orbital time of the gas around the central binary,  $t_{\rm cool} = \beta / \Omega$, with $\beta$ a constant and $\Omega = \sqrt{GM/R^3}$. This
prescription is commonly used in the literature \citep[e.g.,][]{Gammie01, Rice05, NCS07,AACB08}, since the condition for disc fragmentation 
corresponds to a constant $\beta$. As we will discuss later, in a real disc changes to the surface density and opacity with radius result in 
a value of $\beta$ that varies with radius -- we include this physics in our analytic estimates but not in the numerical work.
In this paper we concentrate on the case where the cooling is not fast enough to make the disc fragment, therefore we set the value of
$\beta=10$.  We defer the regime of faster cooling and subsequent star formation, $\beta \simlt 5$, to future study.

The gravity among gas particles is calculated using a Barnes--Hut tree, as implemented in {\sc Gadget-2}.  However, since higher 
accuracy is required when calculating the evolution of the binary orbit we compute the gravitational forces on the SMBHs directly, by 
summing exactly the forces from
all gas particles instead of using a tree approximation. Maintaining symmetry, the gravitational attraction from the SMBHs is added directly to each
particle.

While most of the gas is expected to remain in the disc, accretion streams allow matter to flow from the inner edge of the disc onto the black
holes, where it becomes tightly bound.  To prevent the computation from halting due to the short time-steps required to follow these particles, we
`accrete' all gas that gets within a short distance (a sink radius of $0.1 a_0$) of either black hole \citep{CNSD06}.  The mass and momentum of
the accreted gas are added to the respective black hole, thereby conserving the linear and angular momentum of the system \footnote{This has
been verified {\it a posteriori}.  The change in total momentum of the system is only few$\times 10^{-4} M a_0 \Omega_0$ after
$1000\Omega_0^{-1}$, accurate enough for our purposes.}. When we discuss the `accretion rate' onto the black holes, the reader 
should be aware that it is this accretion of particles that pass within the sink radius that is the actual quantity being 
computed. The physical accretion rate would track the numerical one provided that the residence time of gas through 
the (unmodeled) disc at smaller radii is shorter than the characteristic time scale of accretion variations.

We run each calculation for more than a thousand binary dynamical times using 2 million SPH particles to model the gas.  In addition, to test the numerical convergence of our results, we run shorter simulations with 8 million particles.  

\section{Evolution of the disc}

\begin{figure}
   \centerline{\epsfig{file=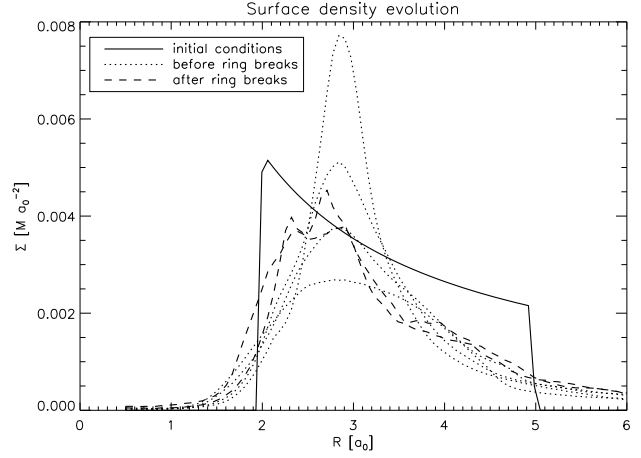,width=.49\textwidth}} 
   \caption{Azimuthally-averaged column density of the disc at times $t \Omega_0 = 0, 50, 150, 250 \ldots 550$.  Initially (solid line), the disc has a surface density profile $\Sigma \propto R^{-1}$ and extends for $2 < R/a_0 < 5$.  Until $t \simlt 350\Omega_0^{-1}$ (dotted lines) the material piles up at $R \approx 3 a_0$ and forms an increasingly dense ring.  This ring finally becomes unstable due to its self-gravity and breaks, making the density profile flatter for $t \simgt 350\Omega_0^{-1}$ (dashed lines).}
   \label{fig:sigma}
\end{figure}

All simulations display very similar disc evolution, so in this section we concentrate on the run with 2 million gas particles and the binary on
an initially circular orbit.  Figures~\ref{fig:disc6} and \ref{fig:sigma} show the evolution of the surface density $\Sigma$ of the disc. 
Figure~\ref{fig:disc6} shows maps of $\Sigma$ at 6 selected times, while Fig.~\ref{fig:sigma} shows the azimuthally-averaged $\Sigma(R)$
profiles at the initial conditions and every $100\Omega_0^{-1}$  in the interval $50 \le t\Omega_0 \le 550$.

\begin{figure}
   \centerline{\epsfig{file=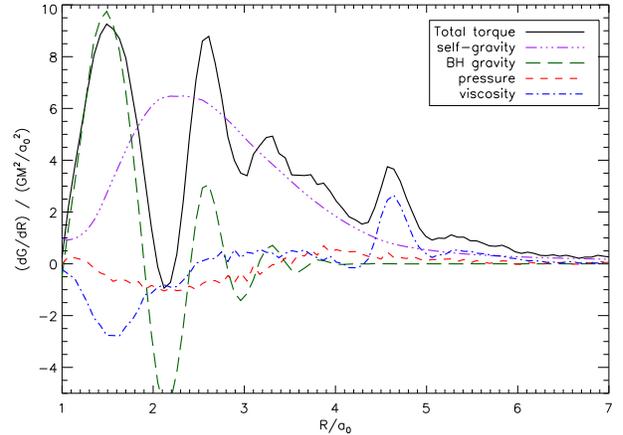,width=.49\textwidth}} 
   \caption{Azimuthally-averaged differential torques $dG/dR$ averaged over the interval $t\Omega_0=$30--50.  Torque is dominated by gravitational effects, both due to the binary (green line) and the disc itself (purple).  Pressure (blue) and artificial viscosity (red) effects are negligible, except at the disc edges, where a very small amount of gas is affected. }
   \label{fig:torques}
\end{figure}

The disc has initially a density profile $\Sigma(R) \propto R^{-1}$, as seen in the solid line of Fig.~\ref{fig:sigma}.  The sharp boundaries
are quickly relaxed and become smoother.  Figure~\ref{fig:torques} shows the different contributions to the torque per unit radius acting at
this early stage in the simulation.  The gravitational potential of the binary produces a torque that oscillates as a function of radius, as
described in detail by \cite{MacFadyen08}.  On top of that there is a large component produced by the disc self-gravity, and smaller
contributions from gas pressure and artificial viscosity.  The total torque has a minimum at $R/a_0 \approx 3$, where material piles up and
eventually a dense ring forms.  There are other minima in $dG/dR(R)$ -- notably one at $R/a_0 \approx 2$ -- but they occur at locations where
not much gas is present, preventing the formation of other density enhancements.  From the figure it is clear that the torques produced by
hydrodynamical forces (both pressure forces, which could be physically important, and artificial viscosity which is a numerical 
concern) are negligible compared with the gravitational torques.  The artificial viscosity torque becomes somewhat important only
at the edges of the disc, due to the fact that shocks form at these locations.

\begin{figure}
   \centerline{\epsfig{file=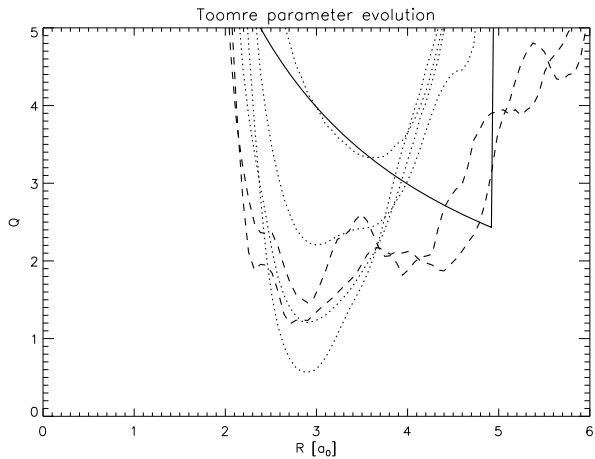,width=.49\textwidth}} 
   \caption{Azimuthally-averaged Toomre parameter of the disc at different times.  The different lines correspond to the same times shown in Fig.~\ref{fig:sigma}.  The disc is initially stable to self-gravity, as $Q > 1$ (solid line), but this value at $R \approx 3 a_0$ decreases with time as gas accumulates in a ring at that location (dotted lines).  After the ring breaks, $Q$ remains larger than unity (dashed lines).}
   \label{fig:toomre}
\end{figure}

As seen in Fig.~\ref{fig:sigma}, the gas piles up at $R \approx 3 a_0$, until the column density becomes so high that the Toomre parameter $Q = \Omega c_{\rm s} / (\pi G \Sigma)$, plotted azimuthally averaged in Fig.\ref{fig:toomre}, becomes smaller than unity and the ring collapses due to its self-gravity.  That process can be see in the panels 2--4 of Fig.~\ref{fig:disc6}, where the ring forms, develops asymmetric structure, and finally collapses after reaching values $Q \approx 0.5$.    The gas that was in the ring then disperses in radius and reaches a quasi-steady state characterised by a complicated spiral pattern and a value of $Q \approx 1.5$.  It is interesting to note that the critical value of $Q$ is different when the gas accumulates in a ring, at the early stage of the simulations, compared to the case where the gas is spread over a larger radial range, later in the evolution of the disc.  This shows that the disc behaviour is not completely determined by its (azimuthally averaged) local properties.  Instead, its stability seems to depend on the properties of the gas at neighbouring annuli and on its azimuthal distribution \citep[see, e.g.,][]{Lodato07}.  

\begin{figure}
   \centerline{\epsfig{file=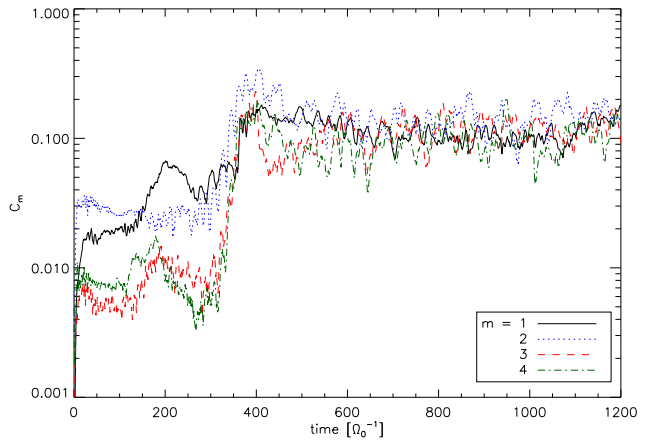,width=.49\textwidth}} 
   \caption{Evolution of the Fourier components of the disc $C_m$ as a function of time.  No mode dominates the structure of the disc after the initial transient period.}
   \label{fig:Cm}
\end{figure}

To characterise the spiral structure that forms in the disc, we measure its Fourier modes, defined as 
\begin{equation}
C_m = \frac{1}{M_{\rm d}} \int_{R_{\rm in}}^{R_{\rm out}}    RdR  \lvert \int_{0}^{2\pi} d\phi \Sigma(R,\phi) \exp(-i m \phi) \rvert   .
\end{equation}
Figure~\ref{fig:Cm} shows the evolution of the $m = 1,2,3,4$ modes as a function of time.  Early in the simulation, during the formation of the ring, only the $m=1,2$ modes are important.  Then all modes grow sharply when the ring breaks after $t \approx 350\Omega_0^{-1}$, but $m=1,2$ remain the most important ones, as the two-armed spiral seen in Fig~\ref{fig:disc6} at $t = 400\Omega_0^{-1}$ suggests.  After a couple of hundred binary dynamical times, however, the disc develops even finer spiral structure, which produces comparable amplitude for all Fourier modes in Fig.~\ref{fig:Cm}. 

\begin{figure}
   \centerline{\epsfig{file=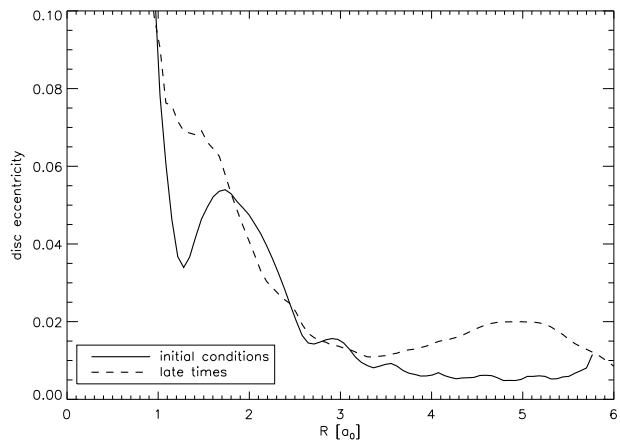,width=.49\textwidth}} 
   \caption{Eccentricity of the disc as a function of radius, both at the beginning and the end of our simulation.  The disc does not develop any significant eccentricity besides that due to the non-Keplerianity of the potential close to the binary, which is already present at the initial conditions.}
   \label{fig:ecc}
\end{figure}

We also measure the eccentricity of the disc, defined as
\begin{equation}
e(R) = \frac{ \lvert \int d\phi \Sigma(R,\phi) v_R \exp(i  \phi) \rvert} {\int \Sigma(R,\phi) v_\phi} .
\end{equation}
Figure~\ref{fig:ecc} shows the eccentricity profile both at the beginning of the simulation, and when we stopped it, 1200 time units later.  Both profiles are built averaging over 20 snapshots, to minimise the random fluctuations.  The initial conditions of the simulation had the gas rotating in circular orbits around the centre of mass of the system.  However, the non-Keplerianity of the potential makes the gas rapidly acquire a small non-zero eccentricity, an effect that is obviously larger closer to the binary.  
After the collapse of the ring, the eccentricity of the disc reaches values $e \approx 0.1$, but then it decays again to smaller values.  As shown in the figure, the eccentricity profile remains approximately the same as initially until the end of the run, $t \approx 1200 \Omega_0^{-1}$, even though the binary reaches $e \approx 0.1$ (Sect.~\ref{sec:orbit}).  In a different run where the binary reached $e\approx 0.3$ the disc did not acquire a significant eccentricity either.  These results are consistent with those of \cite{MacFadyen08}, whose disc only became eccentric after a few thousand binary orbital times. 

\begin{figure}
   \centerline{\epsfig{file=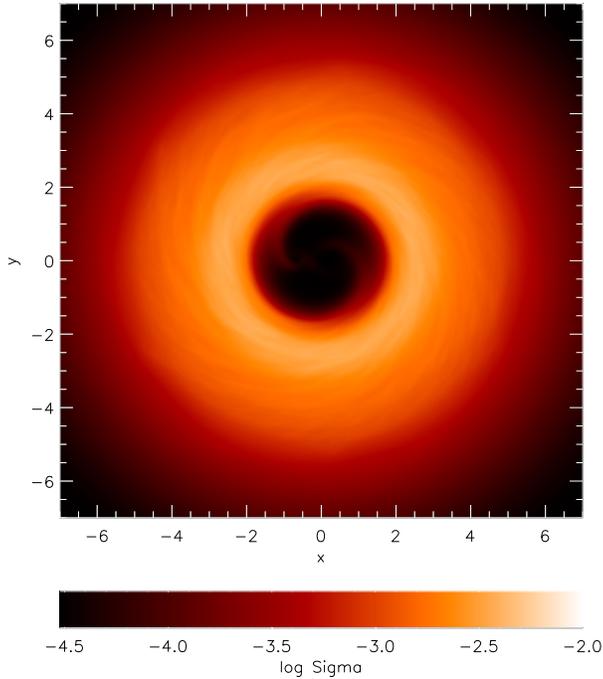,width=.49\textwidth}} 
   \caption{Density map averaged over $t \Omega_0 = 650$--$750$, well after the initial transient evolution, plotted in a frame co-rotating with the binary.  The disc shows no persistent asymmetry.}
   \label{fig:smoothmap}
\end{figure}

To illustrate the long-term evolution of the disc shape, Fig.~\ref{fig:smoothmap} shows a column density map averaging 100 snapshots at $t
\approx 700\Omega_0^{-1}$, in a frame corotating with the binary.  The image shows that the disc develops no persistent asymmetry, except for the
streams taking material from the inner part of the disc to the black holes.

\section{Evolution of the orbit}
\label{sec:orbit}

The binary starts in a circular orbit of radius $a_0$.  However, as the binary exerts a positive total torque on the gaseous disc, it transfers
its angular momentum to it, making the binary shrink. Less obviously, the coupling between the disc and the binary results in the 
excitation of binary eccentricity, even (as in our runs) in the case where the gas disc itself remains approximately circular.

\subsection{Analytic expectations}

The rate of angular momentum transfer between a binary and a surrounding accretion disc can be calculated  analytically \citep{Pringle91}. Most
relevant for our purposes is the work of  \cite{Syer95} and \cite{Ivanov99}, who calculated the evolution of a binary on a circular orbit, under
the effect of an circum-binary $\alpha$ disc.  They neglected the self-gravity of the disc, and assumed that the mass of the secondary black
hole to be much smaller than the primary. Although our conditions violate these assumptions, their calculations can provide us with an order-of-magnitude
estimate of the effect we expect to obtain.

In this section we follow Ivanov et al.\ (1999, their section 4.1) to estimate the time-scale for binary shrinking.  We first need to express
the angular momentum transport in the disc due to self-gravity as a viscosity which is a power-law on surface density and radius, 
\begin{equation}
\nu \propto \Sigma^A R^B.
\end{equation}  
The parameters $A;B$ are determined by the cooling mechanism in the disc.  In our case the cooling time is proportional to the dynamical
time-scale of the disc, $t_{\rm cool} = \beta / \Omega$ (Section~\ref{sec:ics}), which sets the viscosity parameter to $\alpha = 0.4/\beta$, 
\citep[e.g.,][]{Rice05}.  That, together with the condition $Q = Q_0 \approx 1.5$ (Fig.~\ref{fig:toomre}), gives $A=2, B=9/2$, independent of
the values of $\beta$ and $Q_0$.

The other important parameter is 
\begin{equation}
S = \frac{2 \pi a^2 \Sigma_0(a)}{M_2}, 
\end{equation}
which is roughly the mass ratio between the disc and the secondary black hole.  
$\Sigma_0$ is defined as the surface density the disc would have in the absence of the secondary. Over the relatively 
short time scales of the simulations self-gravity results in relatively modest (less than an order of magnitude) 
changes to the disc surface density. To a reasonable approximation we can then extrapolate the $\Sigma(R)$ profile we
use as initial conditions to $R=a$ and get $\Sigma_0(a) \approx 0.01 M a^{-2}$.  That, together with the mass of the secondary, $M_2 = 0.25 M$,
gives $S \approx 1/4 $.

With the parameters obtained above, we follow \cite{Ivanov99} calculation for the evolution of the surface density of a circumbinary disc, under the assumption that the torque of the binary only affects a narrow ring in the inner part of the disc.  We obtain that the time-scale for the black holes to merge is, for the case of a self-gravitating disc,
\begin{eqnarray}
\label{eq:tev}
t_{\rm disc} = \frac{1}{\sqrt{0.93 S}} \frac{M^{3/2}}{0.81 \alpha Q_0^2 \pi^2 G^{1/2} \Sigma_0^2(a_0) a_0^{5/2}} \nonumber \\
= \sqrt{ \frac{M_2}{M} } \frac{1}{19.3 \alpha Q_0^2 } \frac{M^2}{G^{1/2} \Sigma_0^{5/2}(a_0) a_0^{7/2}}.
\end{eqnarray}
Using the parameters of our model, we estimate that $t_{\rm disc} \sim 3\times10^4  \Omega_0^{-1}$.

\subsection{Numerical results}

\begin{figure}
   \centerline{\epsfig{file=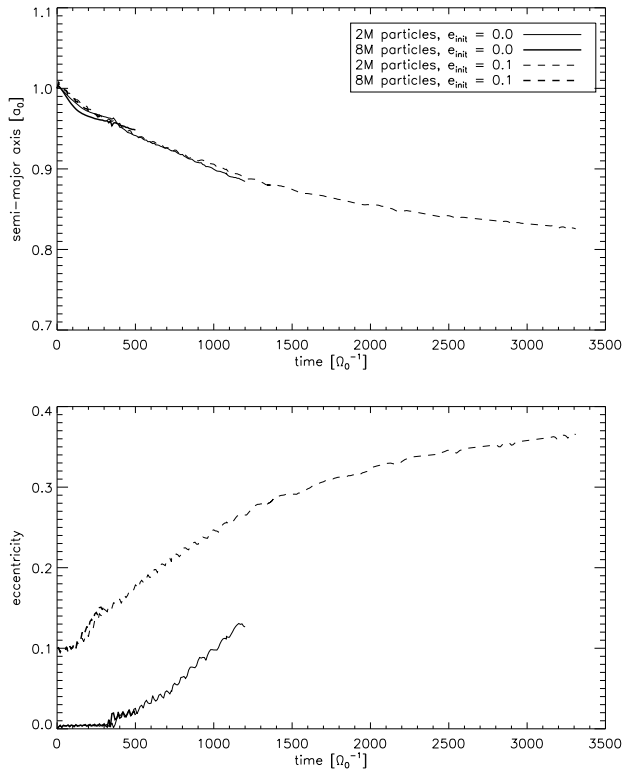,width=.49\textwidth}} 
   \caption{Evolution of the binary orbit elements $a$ (top panel) and $e$ (bottom) for the different simulations.  Solid lines show simulations with the binary in an initially circular orbit, while dashed lines show binaries with $e=0.1$ initially.  In both cases the eccentricity grows, and the orbital decay proceeds at the same rate.  Thicker lines show shorter simulations with higher resolution that agree remarkably with the standard runs.}
   \label{fig:orbits}
\end{figure}

Figure~\ref{fig:orbits} shows the evolution of the binary orbital elements from our simulations.  The orbit of the binary shrinks initially at a rate $da/dt \sim - 10^{-4} a_0 \Omega_0$.  The orbital decay gets somewhat stalled during the accumulation of material in a ring ($t \approx 350 \Omega_0^{-1}$, see Fig.~\ref{fig:disc6}), but it recovers after the ring breaks.  Practically the same rate of change is found for a binary that is initially in an orbit with $e = 0.1$.  That particular simulation was continued until $t \approx 3300 \Omega_0^{-1}$; from that run it is clear that in the long term the rate of decay slows down, reaching $da/dt \sim - 2\times 10^{-5} a_0 \Omega_0$.  We attribute this to the fact that the gas disc continuously absorbs angular momentum from the binary and in average moves farther away from it -- as this process goes on, it is increasingly hard for the gas to influence the binary.  The order of magnitude of the decay rate is in good agreement with the theoretical expectations from the previous section \footnote{Due to the long computing
times required to complete these simulations, we have not yet
explored the parameter space of different mass ratios.  However, preliminary results show that
lower disc masses result in slower binary decays, as expected.}.

\begin{figure}
   \centerline{\epsfig{file=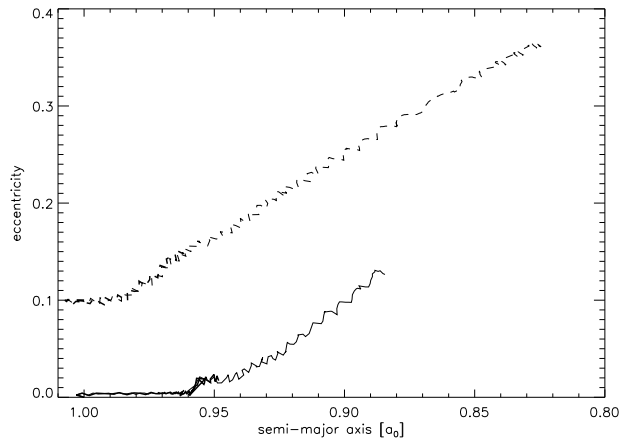,width=.49\textwidth}} 
   \caption{Eccentricity as a function of semi-major axis for the different simulations, as shown in Fig.~\ref{fig:orbits}. 
   Notice that the horizontal axis is inverted, as $a$ decreases with time. The rate $de/da$ remains roughly constant while the evolution of both parameters slows down with time.
   The thick lines show the results of a convergence test conducted with 8 million particles, the light curves refer to longer 
   duration runs using the standard resolution of 2 million particles. The small amplitude variability results from the 
   non-Keplerian potential produced by the disc, which means that the instantaneous Keplerian orbital elements oscillate slightly 
   over the course of each orbit.}
   \label{fig:orbits_ae}
\end{figure}

We also study the evolution of the binary eccentricity.  When the orbit of the binary is initially circular, it remains so during the first part
of the calculation, during which the disc is still mostly symmetric (Fig.~\ref{fig:Cm}).  However, after the disc becomes more asymmetric, the
eccentricity starts growing at a rate $de/dt \sim 1.5 \times 10^{-4} \Omega_0$.  In a different simulation, where the binary orbit is initially
eccentric, the eccentricity growth of the disc starts earlier, at $t \approx 150\Omega_0^{-1}$, but proceeds at approximately the same rate.  As
in the case of the binary size, the eccentricity growth slows down with time.  It is interesting however to study the relation between
eccentricity and binary separation.  Figure~\ref{fig:orbits_ae} shows that $de/da$ remains approximately constant in the long simulation.  We then
conclude that the slower eccentricity growth at later times is mainly produced by the dearth of material to absorb the binary angular momentum. 
It is not due to the onset of any intrinsic damping mechanism, such as might occur if resonant damping grows to balance 
resonant excitation of eccentricity \citep{Moorhead08}. If eccentricity growth saturates in our case, it does so only for 
$e \simgt 0.35$.

The standard resolution of 2 million particles was adopted based on a comparison between the physical torques due 
to self-gravity and numerical torques due to code artificial viscosity. It is not obvious that a resolution that 
minimizes this numerical effect is also adequate to reproduce eccentricity growth accurately. Accordingly, 
we tested the numerical convergence of these results by running shorter simulations with 8
million particles, shown with thick lines in Figs.~\ref{fig:orbits} and \ref{fig:orbits_ae}. These runs give the same results as our 2 million
particle runs.  We are then confident that our simulations capture properly the interaction between the disc and the binary.

\subsection{Maximum rate of orbital decay}
Ultimately, the main question that we wish to address is whether gas discs absorb binary angular momentum 
efficiently enough to drive the black holes toward coalescence on a short time scale. Evidently this will 
depend on the amount of gas that is typically present on very small scales within galactic nuclei. This 
quantity is not well known, either theoretically or observationally. We can, however, estimate the {\em maximum} 
rate of orbital decay that can be driven by a surrounding thin accretion disc. Such a limit can be derived 
because, although more gas results in more rapid orbital decay, the amount of gas in the disc cannot be 
increased arbitrarily. Eventually, a very massive disc will fragment into stars rather than remaining as 
a fluid structure, destroying the hydrodynamic interaction that leads to shrinkage\footnote{The newly formed stars, of course,
could refill the loss cone and thereby help to shrink the binary via stellar dynamical processes. We defer a calculation 
of that process to future work.}.

To estimate the maximum rate of decay we relax the constant $\beta$ cooling time assumption that we introduced 
for numerical convenience. In a real disc the cooling rate will be set by the physical conditions of the disc. 
Some discs will cool faster and fragment into stars, failing to transport angular momentum as a disc. We can then compute 
the maximum decay rate by finding the disc solution that maximizes $| da/dt |$, subject to the constraint that the 
disc is everywhere stable against fragmentation.   We emphasize that the 
resulting disc model -- which is marginally stable against fragmentation at every radius -- is deliberately fine tuned and 
is not what one would expect as the outcome of gas inflow toward the nuclei. It suffices to yield a limit, but that 
limit is unlikely to be exactly realized in nature.

From the analyses of \cite{Syer95} and \cite{Ivanov99}, it is clear that the orbital decay will be faster for discs that
are more massive ($t_{\rm disc} \propto \Sigma^{-5/2}$, see eq.~\ref{eq:tev}). The first step is therefore to 
calculate the maximum surface density, as a function of radius, that a disc can have without fragmenting. We 
follow \cite{Levin07}, and note that the fragmentation boundary in effect represents the maximum stress  
that can be sustained in a quasi-equilibrium self-gravitating disc,  
and can be expressed in terms a maximum allowed value of $\alpha$  
\citep{Rice05}. If in addition the disc opacity is a function of  
temperature only, then there is an analytic solution for a critically  
self-gravitating disc with $Q=1$.  Each value of the  
temperature corresponds to unique values of $\Omega$ and $\Sigma$, and  
together these specify the critical solution. We obtain
\begin{eqnarray}
\label{eq:T}
T = \frac{2 m_{\rm p}}{k_{\rm b}}   (\frac{\pi G \Sigma}{\Omega})^2  ,\\
\label{eq:tau}
\tau = \frac{\kappa(T) \Sigma}{2}  ,\\
\label{eq:alpha}
\alpha = \frac{8 \sigma_{\rm SB}}{9} (\frac{2 m_{\rm p}}{k_{\rm b}} )^4 (\pi G)^6
\frac{\tau}{1 + \tau^2} \Sigma^5 \Omega^{-7},
\end{eqnarray}
where the opacity $\kappa(T)$ is taken from the \cite{Semenov03} opacity tables.  The calculation is stopped when the temperature reaches $\sim 1300\,$K, as the opacity drops sharply there, allowing in principle unreasonably large disc masses.  Moreover, the critical value of $\alpha$ can be much larger in this regime \citep[e.g.,][]{Johnson03}, rendering this analysis invalid.

\begin{figure}
   \centerline{\epsfig{file=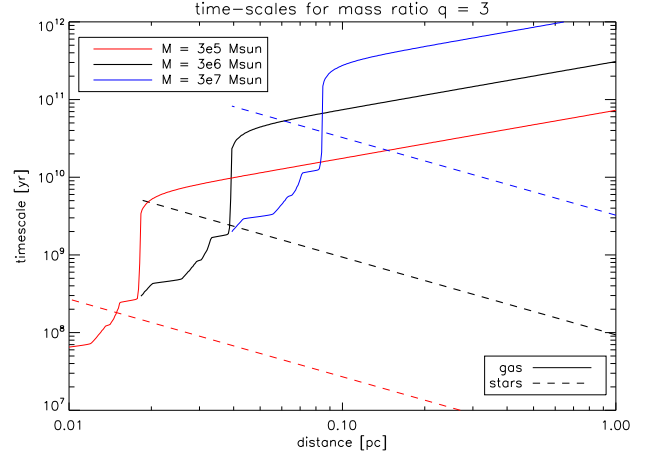,width=.49\textwidth}} 
   \caption{Time-scale for orbital decay for binaries with mass ratio $q=3$ and different total masses.  The solid line shows the maximum effect of a gaseous disc as found in this paper, while the dotted line shows the effect of stellar scattering from \cite{Milosavljevic03}}
   \label{fig:timescales_m}
\end{figure}

\begin{figure}
   \centerline{\epsfig{file=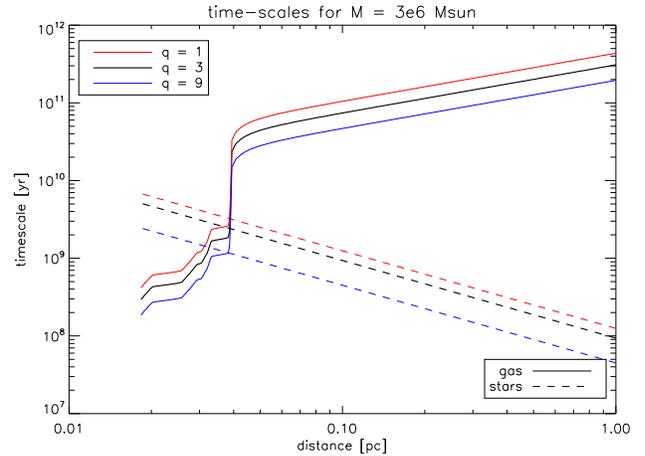,width=.49\textwidth}} 
   \caption{As Fig.\ref{fig:timescales_m}, but for binaries with total mass $M=3\times10^6\msun$ and different mass ratios.}
   \label{fig:timescales_q}
\end{figure}

Once a maximum $\Sigma(R)$ is established, we can calculate the maximum rate of binary shrinking using  eq.~\ref{eq:tev}.  Figures~\ref{fig:timescales_m} and \ref{fig:timescales_q} show in solid lines this time-scale for systems with binary masses $M = 3\times10^5, 3\times10^6, 3\times10^7  \msun$, and mass ratios 1,3,9, as a function of binary separation.  At large separations, the mass needed to make the disc fragment is small, and therefore the removal of angular momentum from the binary is not efficient.  At short separations, much more mass is needed to make the disc fragment, so the process can be much faster.  

The sharp break that appears in every curve is due to the transition from the optically thin regime at large separations to the optically thick one.  At large distances and low temperatures, the  opacity is dominated by ice grains and goes roughly as $\kappa = \kappa_{\rm ice}T^2$, with $\kappa_{\rm ice} \approx 2 \times 10^{-4}\,$cm$^2\,$g$^{-1}\,$K$^{-2}$ \citep[e.g.,][]{Bell94}.  In addition, for a self-gravitating disc the disc properties are linked by $T \propto (\Sigma/\Omega)^2$, which gives an optical depth $\tau \propto \Sigma^5/\Omega^4$.  Looking at eq.~\ref{eq:alpha}, it is clear that the relation between surface density and orbital frequency for a fixed $\alpha$ goes as $\Sigma \propto \Omega^{11/10}$ for optically thin discs, while there is no constraint for optically thick discs, as the $\Sigma$ dependence cancels out.  This explains the sharp decline of the time-scale at the point where the disc becomes optically thick -- arbitrarily large masses are in principle allowed, making the angular momentum transfer through the disc very efficient.  This behaviour, however, breaks down once the gas reaches $T \approx 166\,$K and the opacity law changes to a shallower power-law on temperature.  Further changes of slope are seen in the curves at even shorter separations, which are due to substructure in $\kappa(T)$.

Overall, the time-scale for binary merger becomes shorter as the distance between the black holes is smaller, reaching a value of only $\sim3\times10^8\,$yr for a $3\times10^6\msun$ binary at a separation of $\sim 0.02\,$pc.  The calculation is stopped at this point, as explained above, because the unperturbed disc would reach $T \approx 1300\,$K. 

For comparison we also show in Figs.~\ref{fig:timescales_m} and \ref{fig:timescales_q} the time-scale for orbital decay produced by scattering stars, $t_{\rm stellar}$, as estimated by \cite{Milosavljevic03}, who give 
\begin{equation}
t_{\rm stellar}  \sim 3\times10^9 {\rm yr} \frac{a}{0.1 a_{\rm hard}}  \frac{10^6\msun}{M},
\end{equation} 
where $a_{\rm hard}$ is the binary separation at the point where its orbital velocity is comparable to the galaxy velocity dispersion $\sigma$.  For $\sigma$, we take the value given by the $M$--$\sigma$ correlation, $\sigma \approx 60 {\rm km\,s}^{-1} (M/10^6\msun)^{0.23}$ \citep[e.g.,][]{Ferrarese06}, which yields
\begin{equation}
\label{eq:stellar}
t_{\rm stellar} \sim 5\times10^8 {\rm yr} \frac{q}{(1+q)^2} (\frac{M}{3\times10^6\msun})^{1.54}
	\frac{1 {\rm pc}}{a} .
\end{equation}

\begin{figure}
   \centerline{\epsfig{file=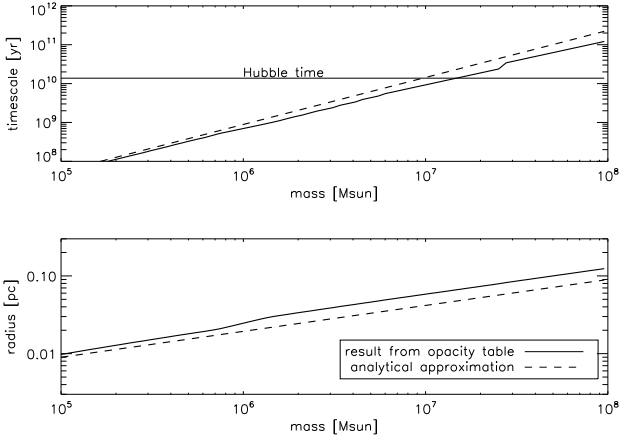,width=.49\textwidth}} 
   \caption{{\it Top: }`Hung-up' time-scale for the evolution of the system including both stellar scattering and the effect of the disc. {\it Bottom: } Binary separation at that point.  Binaries with masses $M \simlt 10^7\msun$ merge in less than a Hubble time.   }
   \label{fig:timescale_m}
\end{figure}

Stellar scattering becomes very inefficient at sub-pc distances from the binary, making the time-scale for decay longer than the Hubble time.  The effect of the disc, instead, becomes more efficient at shorter distances.  Where both time-scales intersect the evolution of the binary is the slowest.  It is important to characterise this `hang-up' point, as the time-scale there will give us an estimate of how fast the binary can merge, and whether it can do it in a Hubble time.  Moreover, we would expect binaries to spend most of their evolution at that distance.  Figure~\ref{fig:timescale_m} shows the values of the hung-up time-scale $t_{\rm hu}$ and separation $a_{\rm hu}$ as a function of binary mass $M$, for mass ratio $q=3$ (there is no big difference in the range $q = 1$--10, as is clear from Fig.~\ref{fig:timescales_q}).  For masses $M \simlt 10^7\msun$, $t_{\rm hu}$ is shorter than the Hubble time, therefore it is possible for such a binary to merge through the method described in this paper.  Such binaries could most likely be caught at distances $a_{\rm hu} \sim$0.01--0.06 pc.

For convenience, we also present an analytical estimate for both $t_{\rm hu}$ and $a_{\rm hu}$.  From Figs.~\ref{fig:timescales_m} and \ref{fig:timescales_q}  it is clear that, for the masses we are interested in, the time-scales for both disc-driven and star-driven evolution intersect roughly where the disc-driven evolution turns due to the optically thin--thick transition.  Numerically, we find that this occurs at $\tau \approx 10$.  Using $\kappa \propto T^2$, as appropriate for this regime, and eqs.~\ref{eq:T}--\ref{eq:alpha}, we find that the condition $\tau = \tau_0$ translates to a fixed value of the binary orbital frequency,
\begin{eqnarray}
\Omega_{\rm hu} = (\frac{4 \tau_0 m_{\rm p} \pi G}{k})^{2/3} \frac{1}{Q_0^{8/3}} (\frac{4 \sigma_{\rm SB}}{9 \alpha \kappa_{\rm ice} })^{1/3} \\
= 7.89\times10^{-10} {\rm s}^{-1} (\frac{\tau_0}{10})^{2/3}(\frac{Q_0}{1.5})^{-8/3}(\frac{\alpha}{0.1})^{1/3} ,
\end{eqnarray}
which corresponds to a period of approx.\ 250 yr. The most common type
of binary would then exhibit phase-dependent variations of the
accretion rate on a time scale that is too long to be readily
observable. Rarer binaries closer to merger would make better
candidates for detection via accretion variations.  A binary of total
mass $M$ will reach the hang-up point at a separation $a_{\rm hu} =
(GM/\Omega_{\rm hu}^2)^{1/3}$,
\begin{eqnarray}
a_{\rm hu}(M) = \frac{M^{1/3}}{G^{1/9}} (\frac{k}{4 \tau_0 \pi m_{\rm p}})^{4/9} Q_0^{16/9} (\frac{9 \alpha \kappa_{\rm ice}}{4 \sigma_{\rm SB}})^{2/9} \\
= 2.77 \times 10^{-2} {\rm pc} (\frac{M}{3\times10^6\msun})^{1/3} (\frac{\tau_0}{10})^{-4/9} (\frac{Q_0}{1.5})^{16/9} (\frac{\alpha}{0.1})^{2/9}. 
\end{eqnarray}
We finally evaluate the stellar scattering time-scale at that position to obtain the hung-up time-scale, 
\begin{eqnarray}
t_{\rm hu}(M) = 1.74\times10^{10} {\rm yr} \times \\
\frac{q}{(1+q^2)} (\frac{M}{3\times10^6\msun})^{1.21} (\frac{\tau_0}{10})^{4/9} (\frac{Q_0}{1.5})^{-16/9} (\frac{\alpha}{0.1})^{-2/9}.
\end{eqnarray}
These analytical approximations are plotted as dashed lines in Fig.~\ref{fig:timescale_m}, and provide a good order-of-magnitude estimate.

\section{Accretion onto the black holes}

\begin{figure}
   \centerline{\epsfig{file=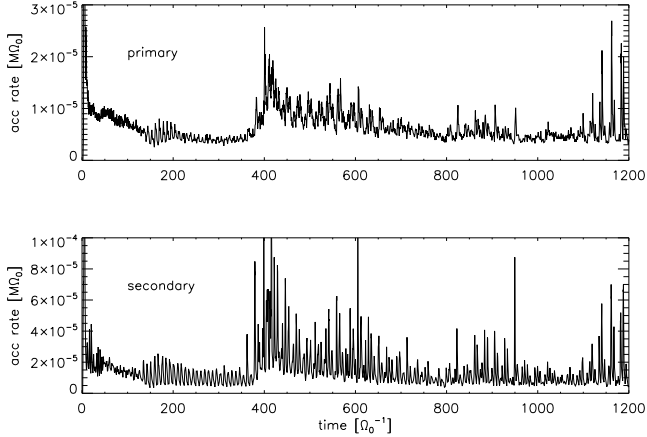,width=.49\textwidth}} 
   \caption{Accretion rate onto each black hole as a function of time.}
   \label{fig:accretion}
\end{figure}

Gas that gets within $0.1 a_0$ of either black hole is taken away from
the simulation to prevent the very short time-steps that would be
necessary to follow it.  This numerical trick is also useful to study
the accretion onto each black hole, in particular its variability and
behaviour as a function of the orbital phase.
Figure~\ref{fig:accretion} shows the accretion rate onto each black
hole as a function of time.  Overall, more gas is accreted by the
secondary black hole, which is closer to the gaseous disc and has a
greater specific angular momentum.  Both curves show a large peak at
the beginning, which is a result of the initial conditions -- gas that
was situated inside the tidal truncation radius.  The accretion rates
increase again at $t \approx 400\Omega_0^{-1}$, after the disc loses
its ring-like shape and a lot of radial mixing occurs.

The accretion rates are strongly variable, especially on the
time-scale of the orbital frequency.  Once a $M \sim 3\times
10^6\msun$ binary reaches a separation of $0.01\,$pc, the orbital
frequency gets to an interesting observable range $\sim
0.1\,$yr$^{-1}$.  Observation of variability on these frequencies will
thus help identify the presence of binaries in such close orbits,
which would otherwise be extremely difficult to detect.  One should
note, however, that the maximal accretion rate that a self-gravitating
disk can support is typically a small fraction of the Eddington
limit\footnote{The mass accretion rate for a critically fragmenting
  disc is $\dot M\sim 3 \pi^3 \alpha Q_0^2 G^2
  (\Sigma(\Omega)/\Omega)^3$.  In the optically thin regime this is
  only $\dot M \sim 10^{-6} \msun\,$yr$^{-1}$, roughly independent of
  the binary mass and separation.  For smaller separations the
  accretion rate can grow larger as the disc becomes optically thick,
  but remains sub-Eddington for the whole range of our study.}.
Additional gas infall to small radii \citep{King07} -- bypassing the
region most vulnerable to star formation -- would be needed to produce
very luminous sources

It should be noticed that the accretion rates are somewhat influenced
by numerical resolution.  With 2 million SPH particles, an accretion
rate of $10^{-5} M \Omega_0$ corresponds to only 100 accreted
particles per orbital dynamical time.  A shorter simulation with 8
million particles indeed shows up to factor 2 discrepancies, however,
the long-term trends and short-term variability are mostly the same.

\section{Discussion}

In this paper we studied the interaction of a super-massive binary black hole with a self-gravitating circumbinary gaseous disc. 
Our focus has been on sub-pc scales, where binary hardening due to stellar dynamical processes is thought to be slow, and 
where efficient radiative cooling makes it likely that the gas will form a geometrically thin accretion structure. Studies of 
galaxy mergers, and indirect evidence from our own Galactic Centre, suggest that gas flows to such small scales may be a 
common feature of galactic nuclei, particularly following mergers. If present, the existence of a gas disc 
around the binary can be a dominant factor in determining whether the binary merges and the time-scale for this process. The 
presence of a disc will influence the binary orbit, as it will transport the angular momentum of the binary outward. At the radii 
of greatest interest the evolution of
such gas discs will be dominated by their self-gravity, which will alter the structure of the disc. Our simulations resolve 
angular momentum transport due to self-gravity physically, without requiring any ad hoc prescription to mimic viscosity.

Our simulations show that the disc, after a transient phase due to the initial conditions, 
develops a quasi-steady spiral pattern that lasts for at least hundreds of dynamical times. 
The binary couples gravitationally to the disc, and loses angular momentum. 
This process makes the orbit of the binary decay at a rate $da/dt \sim - {\rm few} \times 10^{-5} a_0 \Omega_0$ 
(for our simulation parameters), in good agreement with theoretical calculations for non-self-gravitating discs.  
The eccentricity of the binary also increases, at a rate $de/dt \sim 10^{-4} \Omega_0$. 

Due to the computational expense of the simulations, we have not followed the system for long enough to see a 
large change in the orbital parameters. Our
direct numerical experiments show only a modest (20\%) decrease in the binary separation, accompanied by a growth 
of eccentricity to values of around $e \approx 0.3$. The evolution of both $a,e$ appears to
slow down with time, approaching asymptotic values $a \approx 0.8 a_0 ; e \approx 0.4$.  However, we attribute this trend to the fact that the
gaseous disc absorbs angular momentum and moves away from the binary, making the interaction less efficient.  This is consistent with a viscous
time-scale $t_{\rm visc} \sim (R/H)^2 t_{\rm orb} / \alpha \sim 3000\Omega_0^{-1}$ at $R\sim3a_0$, where most of the disc mass is located early
in the simulation. On the other hand, the ratio $de/da$ remains roughly constant throughout our calculations, suggesting that the mechanism for
eccentricity growth has not saturated.  We speculate then that, given enough influx of external material (which could easily occur on the
viscous time-scale described above), the orbit will continue its decay and the eccentricity will keep growing, perhaps reaching the very large
values needed to influence the decay in the relativistic regime.

We also studied the accretion onto the black holes during the orbital decay, and found that it is highly variable, especially on the time-scale
of the orbital frequency.  This variability pattern, if observed in AGN, could help us to identify binaries at sub-parsec separations.

Motivated by the good agreement between the decay rate obtained from our simulations, and that predicted analytically by \cite{Ivanov99}, we 
combine 
their formalism with that of \cite{Levin07} to calculate the maximum decay rate that can be obtained due to a self-gravitating disc.  We find that the decay produced by the
disc can dominate over stellar scattering once the binary separation is 0.01--0.1 pc. The time-scale for decay at the critical 
separation where gas disc and stellar processes are equally important is shorter than the Hubble
time for binary masses $M \simlt 10^7\msun$. This implies that geometrically thin gas discs could provide a solution 
to the `last-parsec' problem for binaries in this range, whereas gas discs are unable to drive mergers within galaxies 
hosting the most massive black holes with masses of $10^8 \msun$ and above. However, even for masses where gas discs 
can in principle result in mergers, the time scales are typically rather long. At low redshift, where the typical 
time scale between major mergers for galaxies hosting $10^7 \msun$ black holes is several Gyr \citep{Volonteri03}, 
we would then expect that many galaxies would host binaries with separations close to the hang-up radius of a few 
hundredths of a parsec. At higher redshift the interval between mergers is much shorter -- less than a Gyr -- and 
interactions between three or more black holes in a single system could not be neglected. 

If we accept that most galactic mergers result in black hole mergers, the results presented here suggest that 
geometrically thin gas discs could in principle drive mergers on a short enough time scale for relatively low 
mass black holes at low redshift. Additional processes, either stellar dynamical or hydrodynamic, are needed 
if very massive binaries are to merge, and to avert ubiquitous 3-body interactions (that might result in 
ejections) at high redshift. One logical alternative that we did not explore
here is that provided by even more massive discs. Such discs will fragment and form stars very close to the 
binary, and if those stars can be scattered by the binary prior to exploding as supernovae they will absorb 
the binary angular momentum almost as efficiently as if the disc had remained fluid. Given slower 
scattering, only low mass stars and stellar remnants will be able to contribute to binary decay. The stellar mass 
function is evidently critical in determining whether such a scenario is viable, as is the stellar dynamics 
of stars formed in a disc close to a (probably eccentric) binary black hole. 

\section*{Acknowledgements}

We thank Volker Springel for his advice on modifying {\sc Gadget-2}, Patricia Ar\'evalo for her help  analysing the
accretion rate variability, and Giuseppe Lodato for pointing us to the work of \cite{Ivanov99}. This research was
supported by NASA under grants NNG04GL01G, NNX07AH08G and NNG05GI92G, and by the NSF under grant AST~0407040.  RDA
acknowledges support from the Netherlands Organisation for Scientific Research (NWO) through VIDI grants 639.042.404
and 639.042.607.  This work was supported in part by the National Science Foundation through 
TeraGrid resources provided by the National Center for Supercomputing Applications under
grants AST070025N and MCA08X017, and utilised the Xeon Linux Cluster (tungsten).

\bibliography{biblio.bib}
\bibliographystyle{mnras}

\label{lastpage}

\end{document}